\documentclass{article}

\usepackage{microtype}
\usepackage{graphicx}
\usepackage{subfigure}
\usepackage{booktabs} 

\usepackage{hyperref}
\usepackage{aas_macros}

\usepackage[accepted]{icml2022}

\usepackage{amsmath}
\usepackage{amssymb}
\usepackage{mathtools}
\usepackage{amsthm}

\newcommand{\logg}{\log g}

\icmltitlerunning{Astroconformer: Inferring Surface Gravity of Stars from Stellar Light Curves with Transformer}

\begin{document}

\twocolumn[
\icmltitle{Astroconformer: Inferring Surface Gravity of Stars\\ from Stellar Light Curves with Transformer}



\icmlsetsymbol{equal}{*}

\begin{icmlauthorlist}
\icmlauthor{Jiashu Pan}{equal,nju}
\icmlauthor{Yuan-Sen Ting}{equal,rsaa,soco}
\icmlauthor{Jie Yu}{mps}
\end{icmlauthorlist}

\icmlaffiliation{nju}{School of Astronomy and Space Science, Nanjing University, Nanjing 210093, China}
\icmlaffiliation{rsaa}{Research School of Astronomy \& Astrophysics, Australian National University, Cotter Rd., Weston, ACT 2611, Australia}
\icmlaffiliation{soco}{School of Computing, Australian National University, Acton, ACT 2601, Australia}
\icmlaffiliation{mps}{Max Planck Institute for Solar System Research, Justus-von-Liebig-Weg 3, 37077 Gttingen, Germany}

\icmlcorrespondingauthor{Jiashu Pan}{jspan@smail.nju.edu.cn}
\icmlcorrespondingauthor{Yuan-Sen Ting}{yuan-sent.ting@anu.edu.au}

\icmlkeywords{Stellar Physics, Self-attention}

\vskip 0.3in
]



\printAffiliationsAndNotice{\icmlEqualContribution} 

\begin{abstract}
We introduce Astroconformer, a Transformer-based model to analyze stellar light curves from the \textit{Kepler} mission. We demonstrate that Astrconformer can robustly infer the stellar surface gravity as a supervised task. Importantly, as Transformer captures long-range information in the time series, it outperforms the state-of-the-art data-driven method in the field, and the critical role of self-attention is proved through ablation experiments. Furthermore, the attention map from Astroconformer exemplifies the long-range correlation information learned by the model, leading to a more interpretable deep learning approach for asteroseismology. Besides data from \textit{Kepler}, we also show that the method can generalize to sparse cadence light curves from the Rubin Observatory, paving the way for the new era of asteroseismology, harnessing information from long-cadence ground-based observations.
\end{abstract}

\section{Introduction}
\label{intro}

If single-epoch data sets drove significant breakthroughs over the last decade, the coming years would undoubtedly herald the golden era of multi-epoch data. The highly successful \textit{Kepler} space satellite \cite{kepler1} has yielded time-series data for hundreds of thousands of stars. And superseding it, the TESS mission \cite{tess} further \textit{Kepler}'s ambition by monitoring stars across the entire sky. And complementing these space satellites are the myriad of ground-based facilities. Notably, the US flagship Rubin observatory \cite{lsst} will start taking data in 2023, whose goal is to scan the sky every few days. As ground-based telescopes generally have a much larger aperture, Rubin will observe much fainter stars than \textit{Kepler} and TESS, thus collecting an unprecedented sample of stellar light curves from the Milky Way's outer halo.

Stellar light curves are invaluable: as the brightnesses of stars vary due to physical processes, such as their oscillations and granulation, the light curves can reveal fundamental properties of stars, including their masses, radii and intrinsic brightnesses, a technique known as asteroseismology. Since the masses and intrinsic brightnesses of stars can then be used to infer precise ages and the distances of them, asteroseismology has been playing a critical role in our quest to unravel the history of the Milky Way.

Despite its central role, the tools to analyze the light curves remain rudimentary to date. For a large part, stellar light curves are often assumed to be a Gaussian process and then characterized by the power spectrum to investigate oscillations and granulation \citep{gaussian1, gaussian2, gaussian3}. However, as solar-like oscillations are stochastically excited and intrinsically damped \cite{damp}, the excitation and damping process of solar-like oscillations is a realization of a Levi process, which contain subtle non-Gaussian signals. Additionally, the long-range granulation, i.e., the convection on stellar surfaces, is also primarily non-Gaussian signals.

Asteroseismology is further limited by the lack of the short-cadence observation to detect high-frequency modes of less evolved stars. In light of this limitation, the community has since seen the potential of applying machine learning to capture lower-frequency information. For example, \citet{Hon} applies convolutional neural networks (CNNs), and \citet{swan} applies a $k$-nearest neighbors method to study asteroseismic data. However, such methods come with their limitations. The $k$-nearest neighbors method fails to make robust inference in the less-populated sample space. The convolution layers of CNNs restrict them to only a local perceptive field. As time-series data often have more long-range correlated information excited at different times, such information is better captured with the attention-based mechanism.

In this study, we will explore the utility of Transformer \cite{attention} to capture long-range information. As a proof of concept, we will apply the method to infer the stellar surface gravity ($\log g$) from stellar light curves.
\begin{figure}
\vskip 0.2in
\begin{center}
\centerline{\includegraphics[width=\columnwidth]{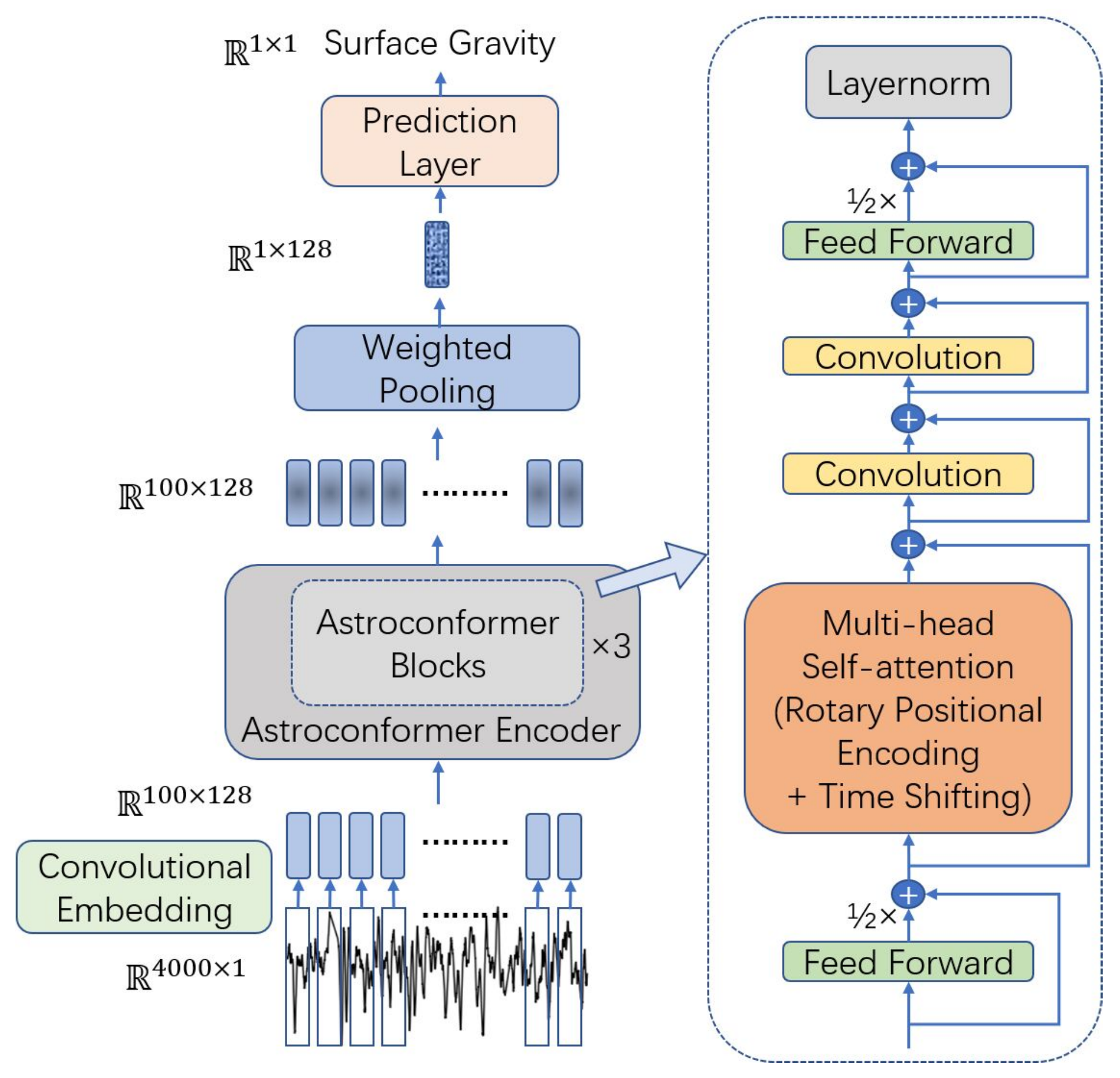}}
\caption{Architecture of Astroconformer. Astrconformer ingests stellar light curves and infer the surface gravity of stars. Our final model comprises three Astroconformer blocks. Each Astroconformer block consists of the convolution and multi-head self-attention layers to capture both the local and long-range information, respectively, going beyond the more explored CNNs method in asteroseismology.}
\label{Astroconformer}
\end{center}
\vskip -0.2in
\end{figure}

\section{Astroconformer}

This study adopts a variant of the Transformer framework known as the Conformer. Conformer was first proposed in the context of speech recognition \cite{conformer}. At its core, Conformer extracts both long-range and local information simultaneously by deploying the self-attention and convolution modules in each ``Conformer block." By interleaving convolutions between Multi-head Self Attention (MHSA) module and the Feed Forward module, Conformer aims to learn the global interaction with MHSA while capturing local features with convolutions. We posit that the Conformer framework is well suited for asteroseismic light surveys because, like speeches, stellar light curves also have correlations on varying timescales.

Fig~\ref{Astroconformer} illustrates our model, which we dub the name Astroconformer. Our method differs from the native implementation in the following aspects. (1) We adopt simply a convolution layer to carry out preliminary feature extraction and tokenize light curves instead of transforming signals into spectrograms as implemented in \citet{conformer}. This is motivated by the fact that the sampling rate of light curves (mins) is much lower than that of speech signals. (2) We include an extra convolution layer in every Astroconformer block. (3) We apply time-shifting technique \cite{timeshift} and rotary positional encoding \cite{RoPE} in MHSA module. (4) After the Astroconformer blocks, we assume weighted pooling across different tokens to adaptively combine and yield the final embedding \cite{weightedpool}.

We will benchmark Astroconformer against the state-of-the-art machine learning method in asteroseismology. In particular, we will compare with {\sc the swan} \cite{swan}, a local linear regression method based on $k$-nearest neighbors. Noting that others also have explored various CNNs to perform a wide range of tasks in asteroseismology \citep[e.g.][]{Hon, cnnrnn}, we will also demonstrate the importance of self-attention by comparing with CNNs through ablation experiments.

\section{Experiments}

\subsection{\textit{Kepler}'s Data and Mock Rubin Observatory Data}

We adopt an identical subset of \textit{Kepler} long-cadence light curves as {\sc the swan}, which facilitates a direct comparison. The data consists of 14003 \mbox{one-quarter} light curves with a median noise of $\sim$ 100 parts per million (ppm). Our training  labels are asteroseismic $\logg$ spanning 0.2-4.3 dex with a typical precision of $\sim$ 0.01 dex \cite{mathur, yu18}. All light curves are normalized by applying a high-pass filter, as was done in \citet{swan}, to remove any low-frequency variations due to the \textit{Kepler} spacecraft.

In addition to the \textit{Kepler} data, we will demonstrate that Astroconformer can be generalized to deal with ground-based observations with a much sparser cadence. In particular, the Rubin Observatory will scan the sky for ten years with a typical cadence of about four days, with an expected noise of 4600 ppm \cite{lsst}. To simulate observations from Rubin, we combine the \textit{Kepler} light curves of the same stars from all \textit{Kepler}'s observed quarters, spanning typically three years in duration. We subsequently degrade the \textit{Kepler} light curves to mimic the noise profile and cadence (assumed to be fixed) of that of the Rubin Observatory.

\begin{figure}[t]
\vskip 0.2in
\begin{center}
\centerline{\includegraphics[width=\columnwidth]{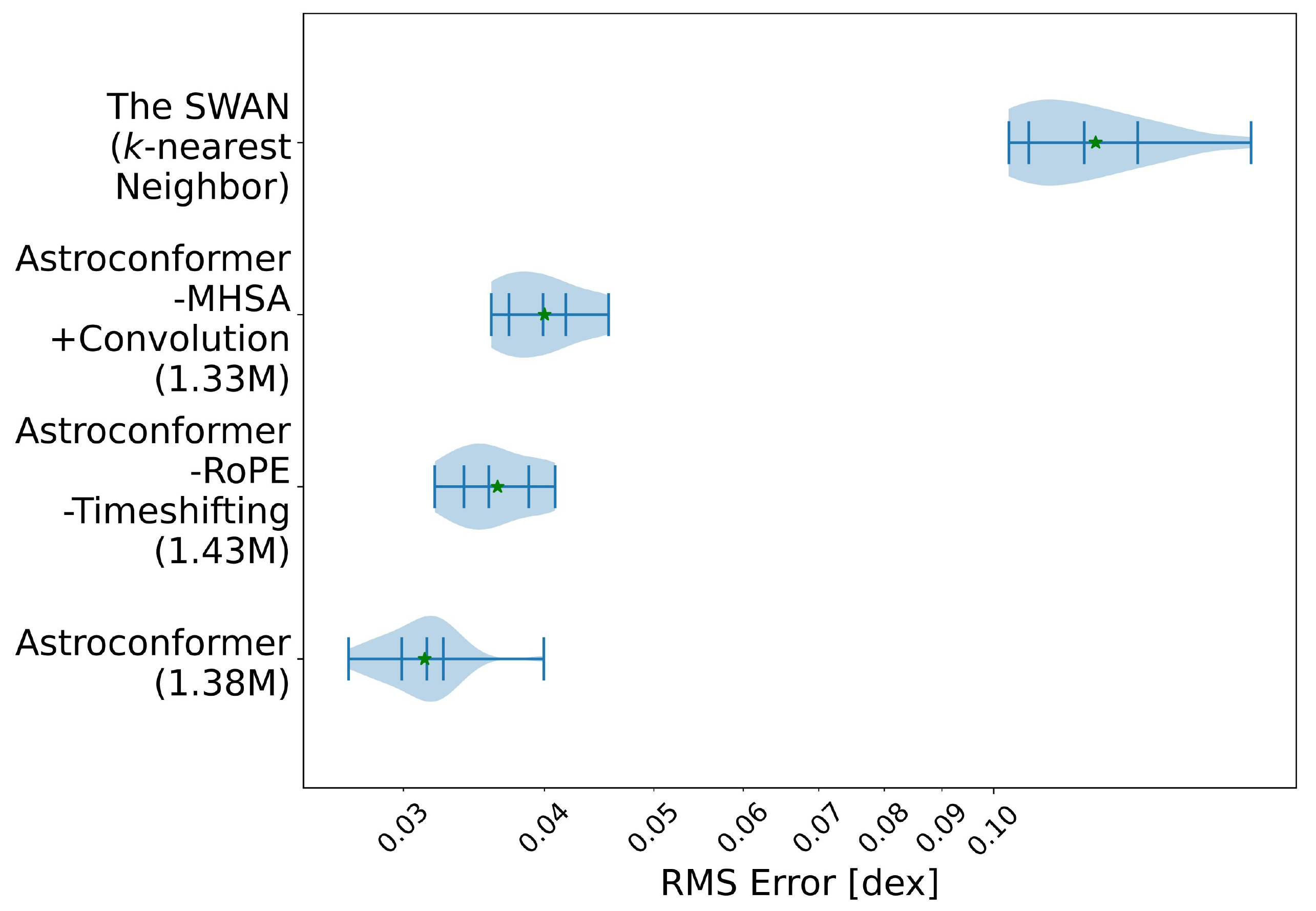}}
\caption{Long-range information harnessed by Astroconformer leads to a more precise inference of the surface gravity of stars. The violin plot shows the RMS errors from the 10-fold cross-validation from the \textit{Kepler} training data and their quantiles. Astroconformer outperforms {\sc the swan} ($k$-nearest neighbors) by threefold and the mere convolution model by about 30\%.}
\label{violin}
\end{center}
\vskip -0.2in
\end{figure}

\begin{figure}[t]
\vskip 0.2in
\begin{center}
\centerline{\includegraphics[width=\columnwidth]{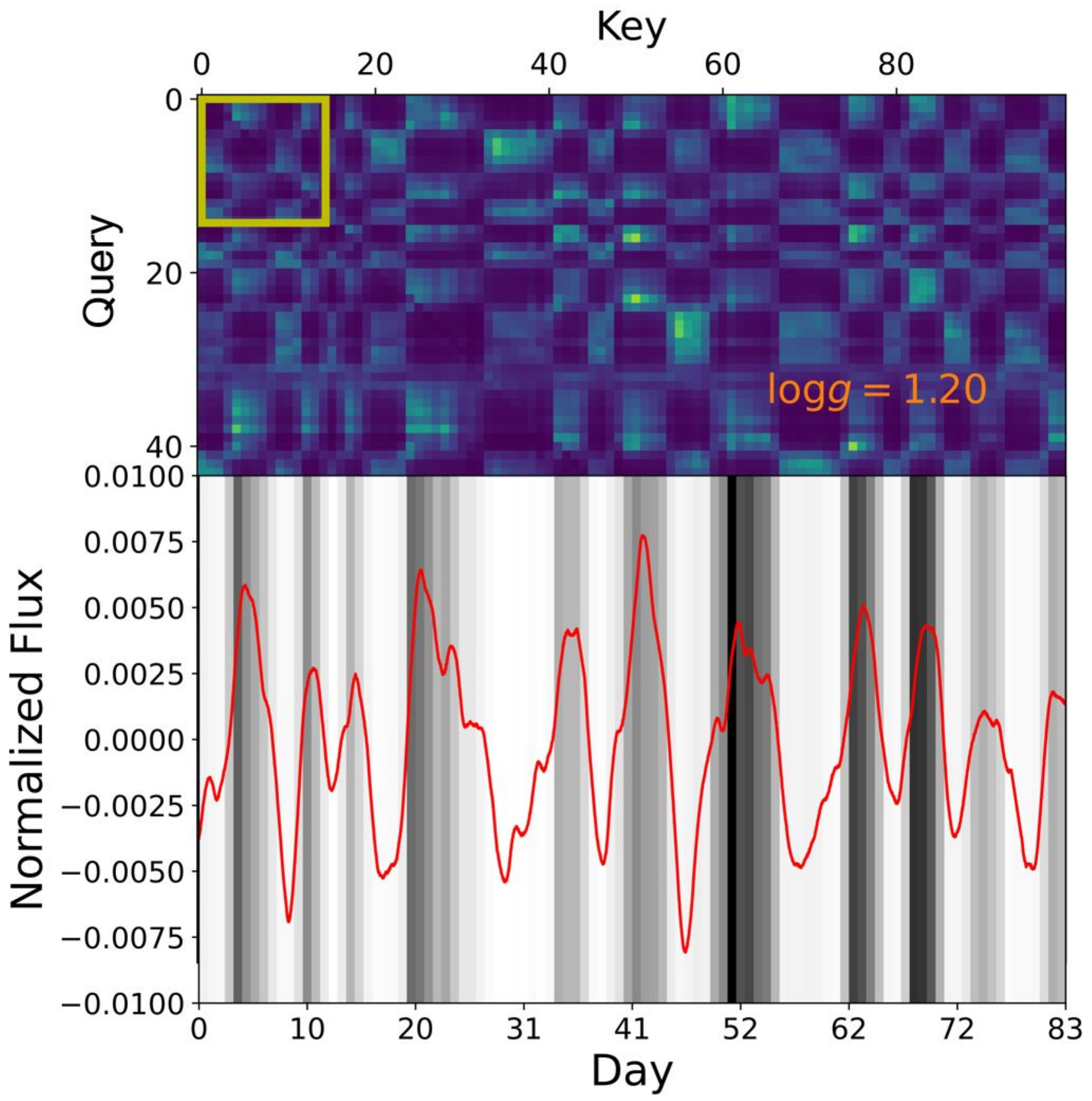}}
\caption{Attentions as learned by Astroconformer. {\it Top:}  Shown is the attention map evaluated for a particular stellar light curve from \textit{Kepler} (KIC 4252170). The yellow square shows the characteristic stellar oscillation period of this star. {\it Bottom:} The light curve with background illustrates the attention value from the first row of the attention map. The attention map shows that Astroconformer extracts not only the characteristic Gaussian oscillations but also more subtle non-Gaussian information within the oscillations.}
\label{attentinomap}
\end{center}
\vskip -0.2in
\end{figure}

\begin{figure}[t]
\vskip 0.2in
\begin{center}
\centerline{\includegraphics[width=\columnwidth]{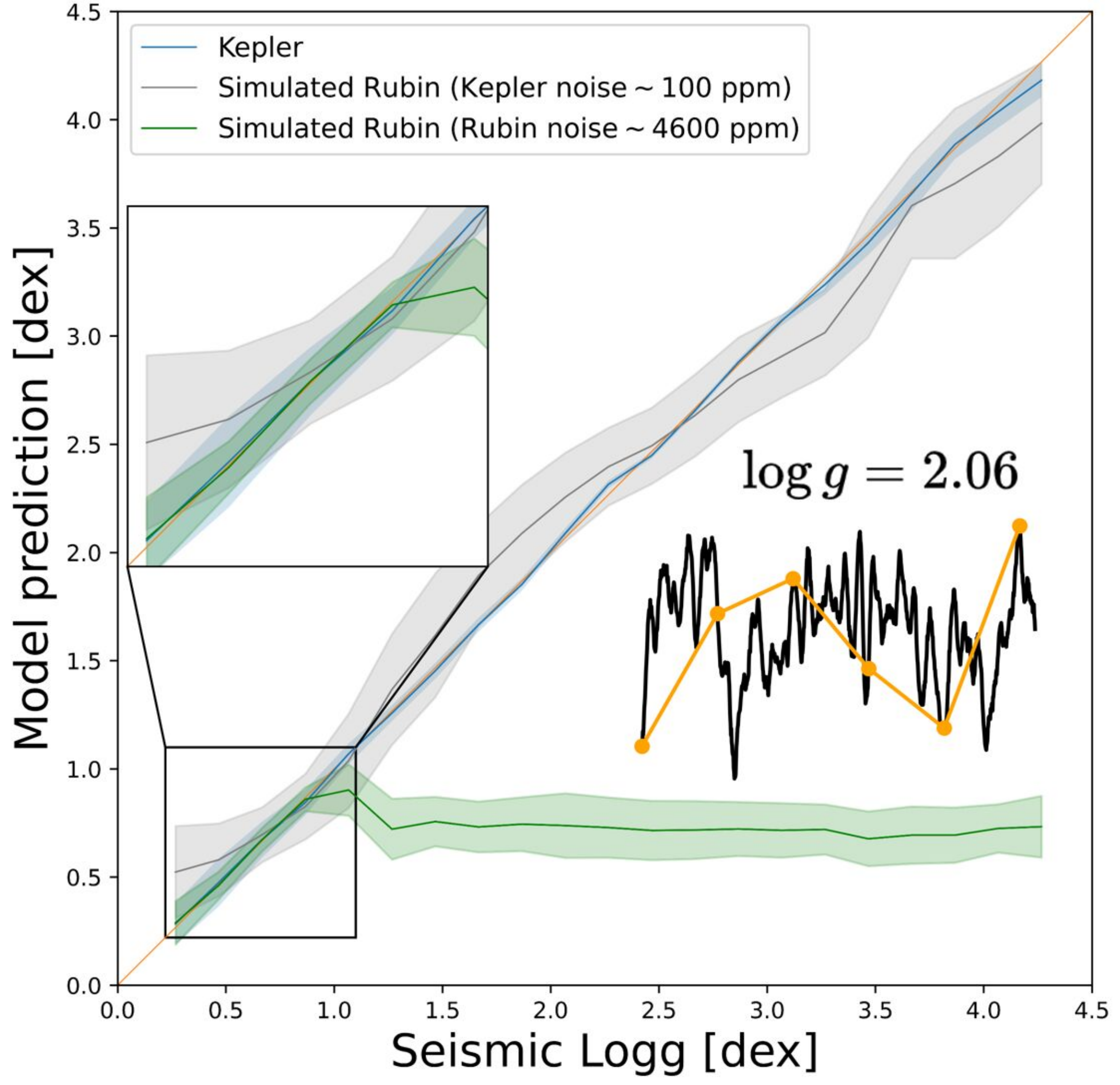}}
\caption{Prospects for inferring surface gravity from Rubin observations. Shown are the 1$\sigma$ error of the $\log g$ inferences on the \textit{Kepler} data (blue) and the simulated data for Rubin (gray $= 100\,$ppm, green $= 4600\,$ppm). Analyzing the much higher cadence of \textit{Kepler}'s data (light curve in black) with Astroconformer leads to an RMS error of 0.027 dex. Although the simulated Rubin Observatory data (light curve in orange) has a cadence of four days, Astroconformer achieves an RMS error of about 0.22 dex with \textit{Kepler}-like noise for all stellar types. With a Rubin-like noise, Astroconformer attains a precision of 0.07 dex for stars with $\log g < 1$.}
\label{Kepler&lsst}
\end{center}
\vskip -0.2in
\end{figure}

\subsection{Performance Comparisons and Ablation Studies}

We first evaluate the performance with our \textit{Kepler} training data to compare Astroconformer with the state-of-the-art method. Then the critical role of self-attention is showed by ablation studies. The ablation studies are tailored to be a fair comparison with approximately the same number of learnable parameters for all models. Evaluations are done by calculating the root mean square (RMS) errors from the 10-fold cross-validation of the training data. The results are summarized in Fig~\ref{violin}.

First of all, and perhaps unsurprisingly, deep learning models outperform {\sc the swan} by a large margin ($\sim$3 folds), with or without MHSA. Our cross-validation studies with \textit{Kepler} data demonstrate that even with only $10^4$ samples in the training data, deep-learning models learn more than just  ``linearly interpolating" the data set, as was done in {\sc the swan}. More importantly, Astroconformer shows that, even with the same training set and the number of learnable parameters, transformer-based model captures long-range information that is critical in characterizing stellar light curves. As shown in Fig~\ref{violin}, removing MHSA from Astroconformer and substituting it with convolution layers leads to the degradation of performance by about 30\% in RMS errors.

On top of that, our ablation experiments also show that, with the still limited training set from \textit{Kepler}, the inductive bias introduced by the bag of tricks in this study (rotary positional encoding and time shifting) has helped the Astroconformer attains better RMS errors ($\sim 20\%$).

\subsection{What has Astroconformer Learned?}

A key advantage of Transformer models, compared to some other deep learning models, is that visualizing the attention can often shed light on what the Transformer models have learned \citep[e.g.][]{bertviz, exBERT}. In Fig~\ref{attentinomap}, we consider a specific \textit{Kepler} light curve and visualize its attention in the first MHSA module.

The attention map illustrates a plethora of information captured by Astroconformer. (1) The attention map presents a periodical block pattern whose period is close to the characteristic oscillation period of the star, signifying the model learns about stellar oscillations. This is further demonstrated in the bottom panel. (2) Importantly, the off-diagonal block shows that our model has understood pixels at varying time stamps are strongly correlated regardless of their location, essentially capturing the long-range information (e.g., granulation) that CNNs miss. (3) Furthermore, Astroconformer also harnesses information at a time scale smaller than the characteristic oscillation, extracting other local ``non-Gaussian" information, presumably from local dynamos that power the stellar oscillations. It is through extracting this combination of Gaussian and local non-Gaussian information and the advantage of capturing long-range correlation through self-attention that explains the superb performance Astroconformer compared to CNNs (see Fig~\ref{violin}).

\subsection{Leveraging Ground-Based Light Curves}

The US-led flagship Rubin Observatory will provide an unprecedented number of stellar light curves from the faintest stars (with magnitude $>$ 26). However, given the ultra sparse cadence ($\sim$4 days) from ground-based observations, to our best knowledge, no one has thus far attempted to infer asteroseismic properties of stars through such light curves. That is because only stars with $\log g < 1$ have oscillation periods longer than $\mathcal{O}(10)$ days. Added to the challenge is the limitation of the power spectrum to determine fundamental stellar properties from uneven, sparse light curves; power spectra have a low-frequency resolution and are subject to the observed cadences that could vary over stars for Rubin.

Without uniform sampling assumption of CNNs, transformer also has the potential to extract information beyond stellar oscillations from unevenly spaced time series \cite{nonuniform}, leading to the possibility of studying the light curves from ground-based missions like Rubin in a meaningful way. To make a preliminary validation of the vision, we attempted Astroconformer for the mock Rubin data which is assumed to be regularly sampled. Fig~\ref{Kepler&lsst} shows that in the limit with the same noise as \textit{Kepler} (100 ppm), even with a cadence of four days, Astroconformer can infer $\log g$ of stars to about 0.22 dex. Since the typical oscillation period of stars for $\log g = 1$ is ten days (near the Nyquist frequency) and even longer for higher $\log g$, extracting only stellar oscillation information would fail. Although here we assume an idealistic noise level, the result demonstrates further that Astroconformer indeed extracts information beyond just the stellar oscillations and can harness subtle non-Gaussian and long-range information even at a cadence of four days.

Nonetheless, with a more realistic noise of 4600 ppm for Rubin, the noise significantly ``Gaussianizes" the signals. The Rubin noise reduces the ability of Astroconformer to most stars with an oscillation period shorter than the Rubin cadence ($\log g > 1$, with a period shorter than 10 days). In this case, we can only infer $\log g$ for stars with $\log g < 1$ to 0.07 dex. Despite that, we emphasize that a 0.07 dex precision for stars with $\log g < 1$ approaches the $\log g$ precision typically attained only by spectroscopy \cite{lamost}. As the surface gravity of giant stars is tightly correlated with the stars' intrinsic brightness, Astroconformer will help refine our measurement of distances from these evolved M-giants, which constitute the bulk of the stellar sample in the Milky Way's outer halo measured by Rubin.

Other considerations also call for further optimism. In this study, we degrade the three-year \textit{Kepler} data to mimic the Rubin Observatory data, but Rubin's 10-year baseline is 2.5 times longer. In addition, Rubin will observe six different photometric bands. The multi-band time series data will contain other subtle information beyond what is in our simplistic mock with degraded monochromatic \textit{Kepler}'s data.

\section{Concluding Remarks}

Astronomy today is a fundamentally different field than it was just a decade ago. The change has been powered by our ability to gather orders of magnitude larger data sets from ever more powerful instruments. But despite this rapid growth in data collection, well-labelled training data remains scarce in astronomy. With the limited training set, implementing neural networks with the relevant inductive bias is critical. As most information from astronomical observations, such as spectra and time series, is long-range by nature, we argue that the local receptive field of the more widely adopted CNNs approach is somewhat limiting.

Our study showed that Astroconformer leverages long-range information and outperforms CNNs in asteroseismology. Since long-range information is prevalent in most astronomical observations, apart from asteroseismology, the same architecture can easily be adapted and might also advance other domains in astronomy. Furthermore, the true power of Transformer lies in unsupervised learning, learning from all the unlabelled data, beyond only those with supervised labels, a subject which we will explore next. Just like the revolution Transformer has sprung across all studies in computer vision and natural language processing, Transformer-based models might prove important to take Astronomy x Machine Learning to the next level.

\bibliography{icml.bib}
\bibliographystyle{icml2022}



\end{document}